# Colloidal graphite/graphene nanostructures using collagen showing enhanced thermal conductivity

Soumya Bhattacharya[1], Purbarun Dhar[2], Sarit K Das[2], Ranjan Ganguly[3] and Suprabha Nayar[1]*

[1]Biomaterials Group, Materials Science and Technology Division, CSIR-National Metallurgical Laboratory, Jamshedpur-831007, India.

[2]Nanofluids, Microfluidics & Bio-MEMS Laboratory, Department of Mechanical Engineering, Indian Institute of Technology - Madras, Chennai-600036, India.

[3]Advanced Materials Research and Applications Laboratory, Department of Power Engineering, Jadavpur University, Kolkata-700098

* E-mail: suprabhan@yahoo.com

**Abstract**

Time kinetics of interaction of natural graphite (GR) to colloidal graphene (G) - collagen (C) nanocomposites was studied at ambient conditions, and observed that just one day at ambient conditions is enough to form colloidal graphene directly from graphite using the protein collagen. Neither controlled temperature and pressure ambience nor sonication was needed for the same; thereby rendering the process "biomimetic". Detailed spectroscopy, X-ray diffraction, electron microscopy as well as fluorescence and luminescence assisted characterization of the colloidal dispersions on day one and day seven reveals graphene and collagen interaction and subsequent rearrangement to form an open structure. Detailed confocal microscopy, in the liquid state, reveals the initial attack at the zig- zag edges of GR, the enhancement of auto-fluorescence and finally the opening up of graphitic stacks of GR to form near transparent G. Atomic Force Microscopy studies prove the existence of both collagen and graphene and the disruption of periodicity at the atomic level. Thermal conductivity of the colloid shows a ~17% enhancement for a volume fraction of less than ~ 0.00005 of G. Time variant increase in thermal conductivity provides qualitative evidence for the transient exfoliation of GR to G. The composite reveals interesting properties that could propel it as a future material for advanced bio-applications including therapeutics.

1. Introduction

The variety of approaches for chemical exfoliation of graphite (GR) is increasing on a daily basis because of the pressing need for bulk synthesis. Till date, the most popular chemical method for GR to graphene (G) conversion is the Hummer's method [1-3]. GR is composed of $sp^2$ bonded carbons within each two-dimensional sheet linked via covalent bonds in an extended array of six-membered rings. But the interactions between sheets are by the much weaker π-bond, i.e., though the G-sheets are individually very strong, in GR, the sheets tend to slide past one another leaving zig-zag edges, close to the Fermi level [4-6]. In addition, there are reports which suggest the presence of sacrificial bonds in the G sheets within GR stacks between adjacent layers and understanding of the chemistry of GR is just beginning to unfold. From a physical viewpoint, chemical exfoliation means that defects can be introduced into GR layers by the attack of chemicals at the above mentioned sites whereby the G sheets crinkle and it becomes relatively easier to break them away from the parent stack. In this work, we report such an attack by aqueous protein dispersion at ambient conditions. The protein used is type I collagen (C) dispersed in acetic acid (A); wherein the C fibrils tend to assemble and exist as self-contained patches because of the "sacrificial bonds" formed by the rupture of non-covalent secondary bonds between the triple helical chain segments in solution and dissipates enormous amounts of energy in the process. Such phenomenon is also observed in many other natural composites like nacre/ bone [7-9]. When C and GR are incubated together, they interact, and with time, form a colloid which seems energetically favorable and the colour of the dispersion changes from light grey to blackish grey (Fig.1). C fibrils can splay discretely between adjacent graphitic layers, disturbing the π−π interactions of GR stacks while simultaneously changing its own tertiary conformation to form the G-C composite. Density functional theory (DFT) calculations and *ab* *initio* molecular dynamics simulations have proved that binding of glycine, proline, and hydroxyproline amino acids of C with G is thermodynamically favorable [10] The reaction has no energy modulation and/or control requirements (temperature, pressure, sonication etc.) like numerous natural processes and hence the process maybe termed "biomimetic". The time-kinetics of the process shows that within 24 hours, the protein attack is complete, after which only internal rearrangements takes place to attain stable equilibrium. This process is safe, overcomes toxicity and enables bulk synthesis of colloidal G; unlike many reports, here C has been used without any modification. Our results are supported by the following developments: i) successful synthesis of a G dispersion from GR flakes in aqueous surfactant solutions by sonication ii) direct exfoliation of GR using ionic liquids having surface tension closely matching the surface energy of GR and iii) "biomimetic composites" based on the exfoliation of GR into a matrix of genetically engineered proteins and nano-fibrillated cellulose [11-14]. No reports of C converting GR to colloidal G at ambient conditions exist till date. This is yet another example of proteins acting as multi-functional macromolecules for biomimetic/bio-inspired functional material synthesis.

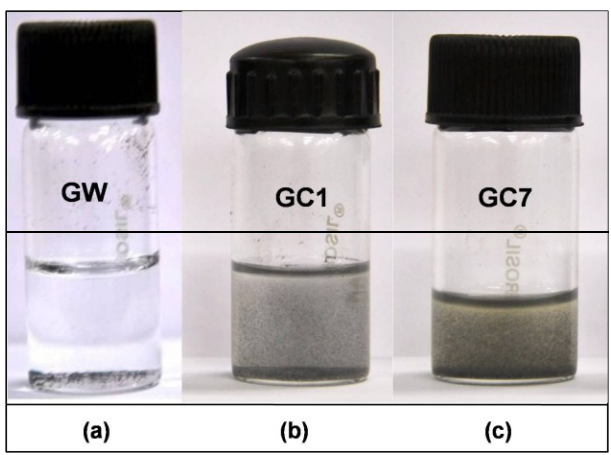

**Figure 1.** GW, GC1 and GC7 dispersions

## 2. Experimental

GR and C was incubated for 1 and 7 days (denoted as GC1 & GC7), after which they were centrifuged at 8000rpm for 30 mins and the supernatant used for all characterization. GR dispersed in water (GW) and in acetic acid (GA) served as controls. Laser Raman spectra were recorded in Nicolet, Almega$^{XR}$ dispersive Raman spectrometer, USA (using a Nd: YAG laser source, $\lambda$ = 532 nm) by drying a drop of the samples on polished copper substrate for surface enhanced signals. X-ray diffraction (XRD) was done in Bruker D8 Discover diffractometer at 40KV with Cu-K$_\alpha$ radiation ($\lambda$=1.5418 Å) in the 2θ range from 6º-35º, step size equal to 0.02º/step and scan speed of 5 secs/step. Hydrodynamic diameter ($D_H$) and polydispersity index (PdI) were measured by Beckman Coulter Delsa$^{TM}$ nano particle analyzer. Transmission electron micrographs (TEM) were recorded in Fei Tecnai G$^2$ T20 s-twin microscope operated at 200kV, all dispersions dried on carbon coated copper grids. PL Photoluminescence (PL) and Fluorescence (FL) were measured in F-4500 Hitachi fluorescence spectrophotometer using the excitation wavelength of 280 nm and emission scan range from 290 - 900 nm for PL and 290 – 450 nm for FL. XPS spectra were recorded in Specs spectrometer with Mg K$_\alpha$ source (1253. 6 eV). Few drops of the dispersions were dried on glass substrates and then inserted into the vacuum chamber of $10^{-9}$ Torr. Spectra were deconvoluted using CASA XPS software, prior to which, Shirley background subtraction was done. All the peaks calibrated w.r.t. the standard C1s binding energy peak of pure GR at 284.5 eV. Atomic force microscopy was done in SPA 400 Seiko microscope in the non contact mode at ambient conditions using pyrex glass cantilever with silicon tip; a drop of the dispersion was dried on glass substrate and the surface imaged. Confocal images were recorded in the fluorescence mode in Zeiss LSM 700 laser scanning microscope using the dyes Alexa Flour 488 (AF488) and Alexa Flour 546 (AF546). No fluorophores were used during the experiments; auto-fluorescence of the samples was measured. A small volume (~2 ml) of the samples was poured in a glass Petri dish, and was directly viewed under the microscope. Thermal Conductivity was measured using the KD2 Pro thermal properties analyzer, Decagon devices Inc. The device works based on the principle of transient hot-wire method. The viscosity was measured using an Anton Paar automated micro-viscometer; which works on the 'falling ball through a capillary' principle.

## 3. Results and Discussion

### 3.1. Surface Enhanced Raman Spectroscopy

Every band in Raman spectroscopy corresponds to a specific vibrational frequency of a bond within the molecule. Characteristic GR peaks (spectra shown in the inset and characteristic peak positions in Table 1) at 1592 cm$^{-1}$ (G band: because of in-plane phonon vibrations of sp$^2$ hybridized carbon atoms in the aromatic carbon rings) and 2728 cm$^{-1}$ (2D band: from a two-phonon double resonance) were seen, along with minor vibrational modes at 2447 and 3250 cm$^{-1}$ [16]. No D band (~1350 cm$^{-1}$) was detected. This band is normally assigned to the vibrations of sp$^3$ hybridized carbon atoms of disordered GR arising due to structural defects, indicating that the precursor GR has minimal defects. As expected, there was no appreciable change in the peak position of the controls GW and GA w.r.t. GR. After incubation with C, we could observe significant changes in the GR peaks which are the following: i) The D peak (1358 cm$^{-1}$) not present in GR, appears in GC1 and increases in GC7, due to the introduction of more and more defects with time in the almost defect free graphitic domains. It has been reported previously that the D peak intensity is very weak in single layer G but increases progressively with the number of layers which indicates the degree of edge chirality [17, 18] ii) the G band (1589, slight shift from 1592 cm$^{-1}$) is present in both GC1 and GC7, the intensity did not differ much but the peaks were more broadened, due to the enhanced interaction with C molecules with time leading to an increase in the opening up of more graphitic stacks. The single sharp peak at 1589 cm$^{-1}$ splits into 3 peaks positioned at 1630, 1589 and 1551 cm$^{-1}$, iii) the 2D band of GR decreases and broadens in GC1 and disappears completely in GC7. Very few research groups have reported the complete disappearance of the 2D peak, the most noteworthy being the paper on fluorographene [19]. $I_D/I_G$ ratio increases from 0 in GR to 0.27 in GC1, increasing further to 0.99 in GC7, a clear proof that with time C and GR are internally rearranging themselves to arrive at an equilibrium state and form stable colloidal G (Fig. 2)

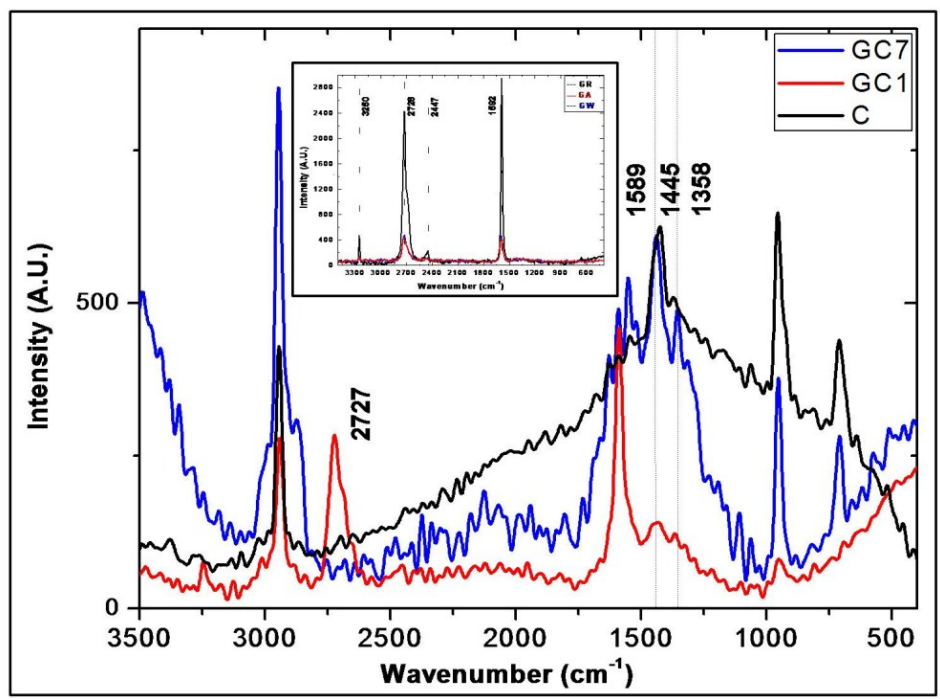

**Figure 2.** SERS spectra of C, GC1 and GC7. The 2D (2727cm$^{-1}$) peak in GC1 vanishes in GC7 signifying interaction. The control spectra are shown in the inset

| SAMPLE | D (cm$^{-1}$) | G (cm$^{-1}$) | 2D (cm$^{-1}$) | | | $I_D/I_G$ | $I_{2D}/I_G$ |
|---|---|---|---|---|---|---|---|
| GR | - | 1586 | 2447 | 2722 | 3251 | 0 | 0.82 |
| GC1 | 1357 | 1589 | - | 2724 | - | 0.27 | 0.62 |
| GC7 | 1357 | 1589 | - | - | - | 0.99 | 0 |

**Common protein peaks for GC1 and GC7: 714, 954, 1424, 2946 in cm$^{-1}$, identified w.r.t. the control spectra C.**

**Table1.** Detailed Raman peak positions of the analyzed samples corresponding to D, G and 2D bands along with the intensity ratios ($I_D/I_G$ & $I_{2D}/I_G$) signifying interaction

*3.2. X-Ray Diffraction*

XRD patterns of C, GC1 and GC7 are shown in Fig. 3, the control spectra of GR, GW and GA have been plotted separately in the inset. All the patterns indexed w.r.t. standard ICDD data for crystalline carbon (JCPDS Card No. 41-1487). C did not show any peak, just a broad hump typical of an amorphous material. GC1 and GC7 exhibited peaks at the positions 26.598° (interplanar spacing ($d_{hkl}$) = 3.348 Å, full width at half maximum (FWHM) = 0.018) and 26.658° ($d_{hkl}$ = 3.341 Å, FWHM = 0.015) respectively. This peak is detected in all the samples corresponding to (002) Bragg reflection of GR. In case of GA, some minor peaks were also detected at other 2θ values (15.36°, 19.39°, 23.90° and 30.93°), which probably appear as a result of acetate crystallization. Ideally single layer G is not expected to show any diffraction peak, the peak increases in intensity with increase in layer thickness. In our experimental data, the intensity profile of GC1 is 3 times lower than that of GC7. As the concentration of GR in both GC1 and GC7 dispersions is the same, such a result can only be explained as the few layer GC1 interacting with each other with time in GC7. The increase in the average crystallite size also indicates such a possibility, the protein functionalized G surfaces are attracted to each other, the layers aggregate but do not re-stack, the resulting stable dispersion consist of a homogeneous composite. Detailed analysis of (002) peak position of all samples is tabulated in Table 2, the average crystallite size corresponding to the (002) peak was calculated using the Scherrer equation [20]. The reported values of interlayer spacing in G are 3.4Å and for bulk GR it is within 3.348 – 3.360 Å, a difference of only 0.052Å [21, 22]. It seems that even after one day the C molecules interact substantially with GR. This is remarkable since no external energy (chemical, physical or mechanical) is supplied to overcome the π-stacking, just molecular rearrangement and mutual energy balance results in this interaction, which increases with time.

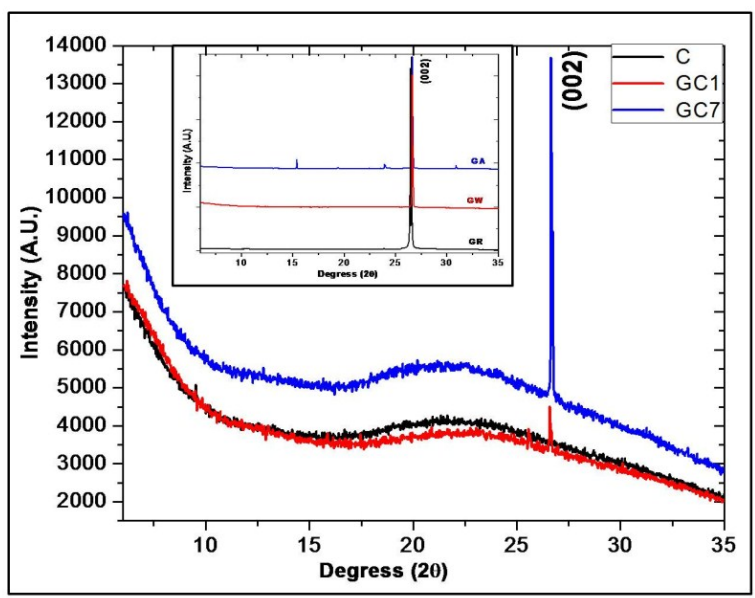

**Figure 3.** XRD peaks of C, GC1 and GC7, the controls shown in the inset. Highest intensity peak was recorded at 26.67° corresponding to (002) reflection

| Sample | Position (2θ, Degress) | $d_{hkl}$ (Å) | FWHM (2θ, Degrees) | Crystallite Size (nm) |
|---|---|---|---|---|
| GW | 26.640 | 3.343 | 0.051 | 157.9 |
| GA | 28.629 | 3.344 | 0.053 | 151.6 |
| GC1 | 26.598 | 3.348 | 0.018 | 427.5 |
| GC7 | 26.658 | 3.341 | 0.015 | 519.7 |

**Table 2.** XRD parameters that have undergone change as a result of C interaction

*3.3. Transmission Electron Microscopy and Dynamic Light Scattering*

TEM micrographs of GC7 (Fig. 4), show the seamless co-existence of C-GR in GC7 (GC1 TEM was not done). The protein molecules seem to have intercalated the graphitic layers and broken the π- stacking. After the initial attack at the zig-zag edges, C-GR interaction has to be by inductive charge destabilization resulting in the disturbance of the π-π stacking of the layers in GR, a phenomenon dictated by the similarity of the interacting molecular structures. With time, the interaction attains a steady state, the graphitic planes and C exist in complete harmony, and thickness contours of the lateral graphitic edges are of the order of 3 nm, shown in Fig. 4 (c-d). The images also clearly show that the point of attack is at the GR edges, unevenly distributed throughout the dispersed phase. The selected area electron diffraction (SAED) patterns (not shown) was different from that of standard G, indirectly proving the role of C. Dynamic Light scattering results (not shown) are complimentary to the TEM data. The hydrodynamic diameter ($D_H$) of C dispersion was 536.6 nm and polydispersity index (PdI) equal to 0.25. In case of GC1 the PdI increased to 0.32, $D_H$ decreased to 382.7 nm, meaning that even after one day in GC1, the graphitic surfaces interact with C and form compact structures. As the ratio of C is 50 times that of GR, GR is presumably breaking down into smaller GR flakes with C attack. After 7 days in GC7, the interactions between C and GR stabilize and the broken smaller GR stacks tend to self assemble and stabilize into a more open system, due to which the $D_H$ again increases to 504.7 nm but PdI decreases to 0.21, an indication that the system is tending towards more stability.

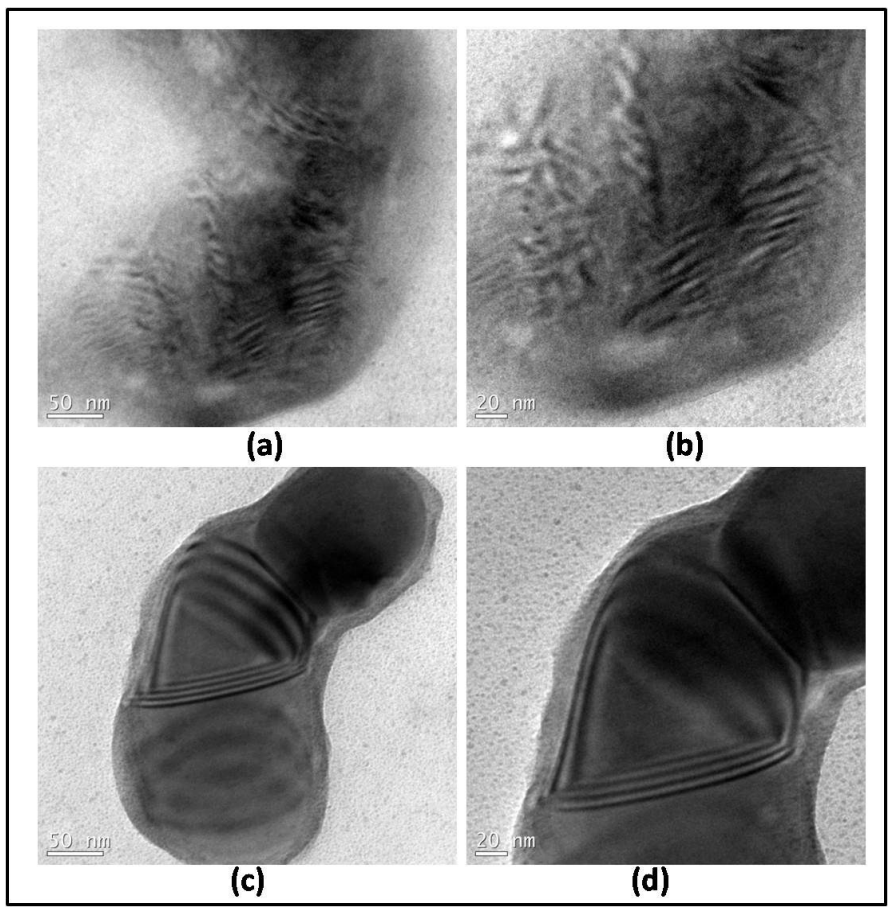

**Figure 4.** TEM micrographs of GC 7 dispersions (a) - (d) showing the seamless co-existence of C and G

*3.4. Fluorescence Spectroscopy*

FL and PL spectra are shown in Fig. 5(a-b). Intrinsic fluorescence (FL) in proteins principally assesses the protein conformation. When the tertiary structure of the protein begins to open up, the interior is more accessible to reagents and this often relates to an increase in instability. Typically, emission maxima below 336 nm indicate that a well folded tertiary structure exists for a protein whilst emission maxima ≥ 340 nm ($\lambda_{max}$) represents significant opening-up of the structure. An increase in FL of GC1 must be because of the contribution from the aromatic structures of GR which also contribute to the enhanced FL, which with time decreases and a new peak appears at around 399nm possibly because of charge transfer as a result of the C-GR interaction. π-π stacking interactions and hydrophobicity drives the charge transfers between C and GR. This is further supported by the photo luminescence (PL) data which increases with openness of the structure, while FL decreases with openness. The PL in GC7 increases dramatically, forming in all probability a sheet like G-C composite. While single/ bilayer G does not exhibit PL, it is also true that the PL strength decreases considerably in thicker sheets, indicating that the thickness of the composite synthesized is thin enough to exhibit PL [23]. This optical property of disordered carbon thin films containing a mixture of $sp^2$ and $sp^3$ carbon is not new. The PL in such carbon systems is a consequence of recombination of localized electron–hole pairs in $sp^2$ clusters, which essentially behave as the luminescence centers and varies depending on the size, shape, and fraction of the $sp^2$ domains. PL energy scales linearly with the $sp^2$ fraction in disordered carbon systems, implying that the GR stacks open up time [24, 25].

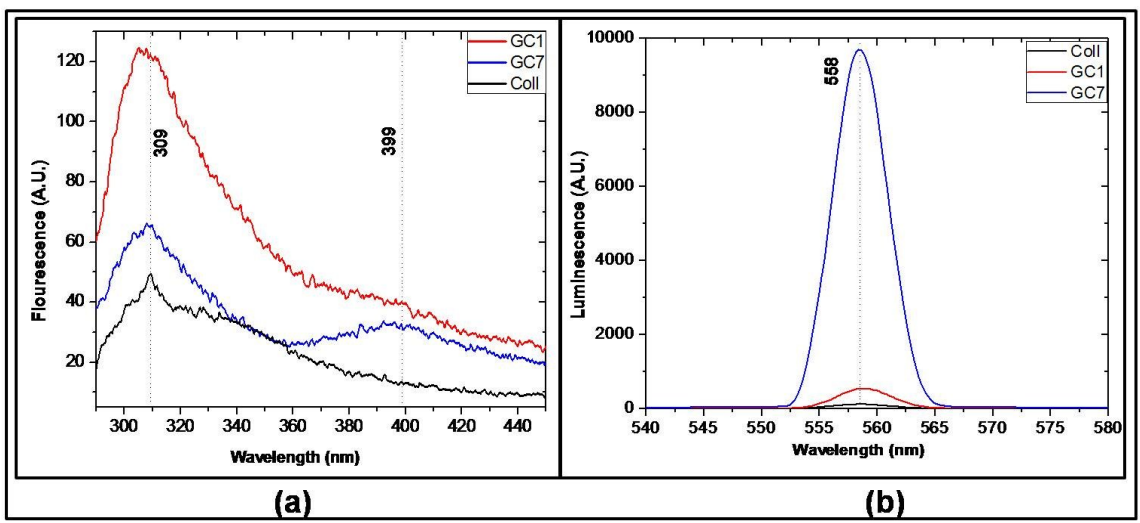

**Figure 5. (a)** FL spectra of C, GC1 and GC7 showing conformational changes in the protein structure (excitation wavelength ~ 280nm, emission range ~ 290-450 nm) **(b)** PL spectra of the samples proving opening up of graphitic stacks (excitation wavelength ~ 280nm, emission range ~ 290-450 nm). The decrease in FL and increase in PL correspond well with the gradual opening of the system

*3.5. X-ray Photoelectron Spectroscopy*

A list of the spectral parameters that ascertains the chemical environment around the carbon atoms of GC1 and GC7 has been tabulated and shown in Table 3. The full spectra (Fig. 6) with binding energy (B.E) ranging from 1000 - 0 eV clearly shows that the main chemical components in GR is C1s (~285eV) and O1s (~534eV); all the remaining samples show the expected N1s (~ 401 eV) also. The GR C1s can be de-convoluted to three peaks with binding energies at 283.101, 284.624 and 285.174 eV, the first, not a standard G peak, may be assigned to impurities present in the precursor GR, the second corresponds to $sp^2$ hybridized C-C domains and the third to $sp^3$ hybridized C=C domains [26-28].

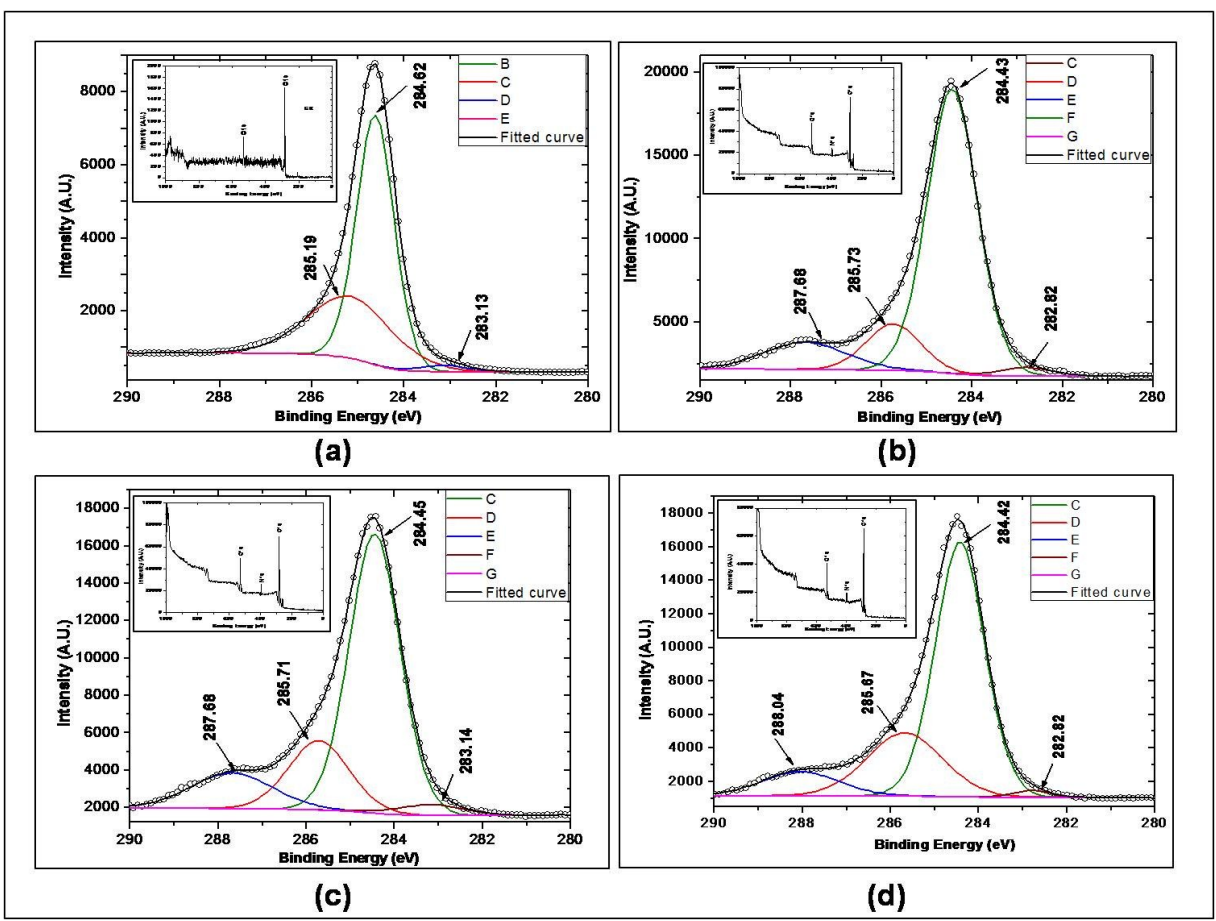

**Figure 6.** XPS spectra of (a) GR, (b) C, (c) GC1 and (d) GC7 dispersions suggesting interaction between C and GR.

All other samples showed de-convoluted peaks at similar B.E., along with the expected additional peak within the 287-288 eV range corresponding to C-NH$_2$ bonds because of the presence of C. This is the position of the p$_z$ orbital and hence it wouldn't be wrong to say that C is finding entry into the π-π* region. Since C is dispersed in acetic acid, several functional species may be present within the same B.E. values *e.g.* >C=O (~286.45-287.92 eV), C-O (~286.21-287.53 eV) and C-NH$_2$ (~286-288.5 eV). The peak width of sp$^2$ C-C domains of GR (0.9585eV, Δ=0eV) increases in GC1 (1.352eV, Δ=0.3935eV) and GC7 (1.318eV, Δ=0.3595eV), the percentage concentration also follows the same trend, GR (62.61, Δ=0), GC1 (64.19, Δ=1.58) and GC7 (64.34, Δ=1.73). Now if we consider the $sp^3$ C-C domains we can clearly see that the peak width increases in GC7 (2.012eV, Δ=0.39) w.r.t. GC1 (1.622 eV, Δ=0), the percentage concentration changes as GC7 (24.50, Δ= 5.41) and GC1 (19.09, Δ=0). This indicates that with time the $sp^3$ character increases meaning more functionalization with proteins. The most striking feature of the XPS spectra is the change in the spectral parameters corresponding to the C-$NH_2$ peak. The percentage concentration increased from C (11.60) to GC1 (13.36) and then decreased in GC7 (9.73), re-confirming our belief of internal rearrangements leading to a more homogeneous composite. The GC1 sheets must be within nanometer range of each other for the effect of edge atoms to become pronounced and for the edges to interact with each other to form an open sheet. The broadening remaining more or less unaffected, 2.162 eV in C, 2.195eV in GC1 and 2.068eV in GC7 meaning that as predicted, the protein attack at graphitic sheets is complete after the first day of incubation. A further observation is the intensity (counts per second, cps) of the π-π* satellite peak of the four samples at 290 eV, it is the lowest for the GR (857 cps) and the highest for C (2245 cps), with GC1 (2062cps) and GC7 (1288 cps) in between, further supporting our observation of $sp^2$ - $sp^3$ transition.

| SPECTRAL PARAMETERS | SAMPLES | | | |
|---|---|---|---|---|
| | GR | C | GC1 | GC7 |
| **Binding Energy (eV)** | 285.174, 284.624, 283.101 | 287.682, 285.716, 284.433, 282.823 | 287.687, 285.716, 284.448, 283.147 | 288.042, 285.676, 284.425, 282.822 |
| **Peak width** | 2.170, 0.9585 | 2.162, 1.422, | 2.195, 1.622, | 2.068, 2.012, |

| (FWHM, eV) | 1.323 | 1.305, 1.387 | 1.352, 1.790 | 1.318, 1.048 |
|---|---|---|---|---|
| % conc. | 35.16, 62.61, 2.24 | 11.60, 13.00, 73.04, 2.36 | 13.36, 19.09, 64.19, 3.36 | 9.73, 24.50, 64.34, 1.42 |

**Table 3.** Spectral parameters showing B.E., peak width and % concentration of deconvoluted peaks corresponding to the C1s spectra of the samples, signifying different functional species bonded with parental carbon.

*3.6. Atomic force microscopy*

Atomic force microscopy (AFM) is a powerful tool in the family of techniques dedicated to nanoscale surface characterization [Ref]. All images shown in figure 7A were recorded in the non contact mode (an AFM technique also known as dynamic force microscopy or DFM) from a particular region on the surface at the same magnification in different modes. Figure 7A (a) gives the surface topography by measuring the atomic force due to tip surface interaction, figure 7A (b) measures the different frictional forces (in millivolt) experienced by the tip as it moves over the sample surface; the image contrast and surface contours arise because of C and G and figure 7 (c) the difference in Van der Waals deflection of the tip while scanning the surface. This is a direct evidence of the co-existence of the carbon compound and protein. Figure 7A (d) gives a three dimensional topographical image of surface atoms, the periodicity of the atomic arrangement is broadened due to the co-existence of the protein matrix (C) and the dispersed phase (G); from the z height of the peaks (~2 x $10^{-2}$ nm) and considering the spring constant of the atomic probe or tip equal to 0.08N/m, the force experienced tip was equal to 0.16 X $10^{-11}$ N, typically within the expected range. The same area was imaged at higher resolution (focus area = 2nm x 2nm); a quantitative assessment of the bond distance between adjacent atoms from two different localized regions have been tabulated. The scanned profile taken along the small line drawn in figure 7B (a) matched exactly with that of the C-C bond distance in graphene (0.142nm); the region looked like a distorted hexagon with an average roughness parameter equal to $2.583 \times 10^{-3}$ nm. However subsequent profile analysis shown in figure 7B (b) had smaller inter-atomic dimensions due to C-G interaction as AFM images are affected by the cross talk between the x, y and z axes; the average roughness increased to $4.721 \times 10^{-3}$ nm indicating the existence of the collagen matrix.

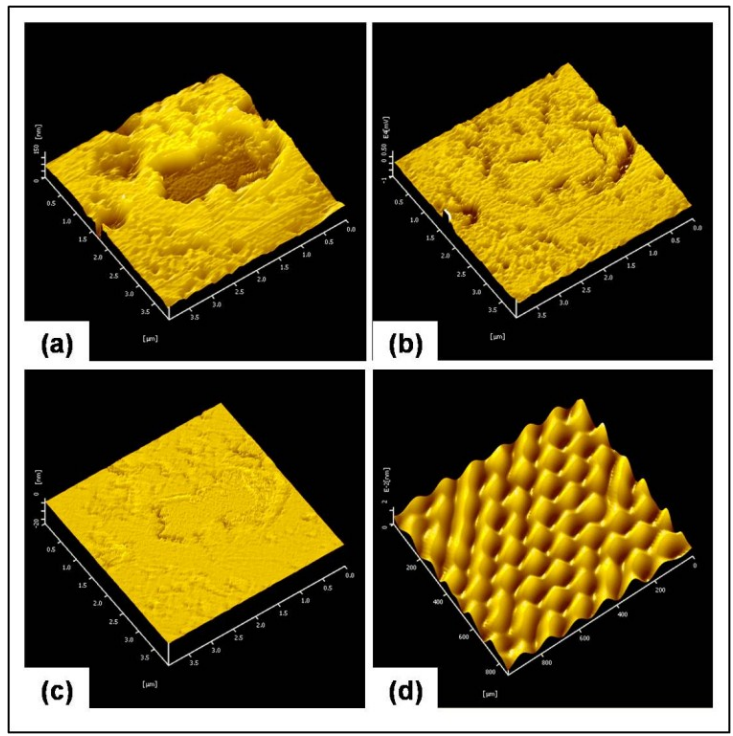

**Figure 7A.** Non-contact AFM images of GC composites on glass (focus area equal to 4μm x 4μm) in different modes (a) topography (b) frictional force and (c) deflection, (d) a topographical image of the surface showing a bigger area of focus 1000 μm x 1000 μm.

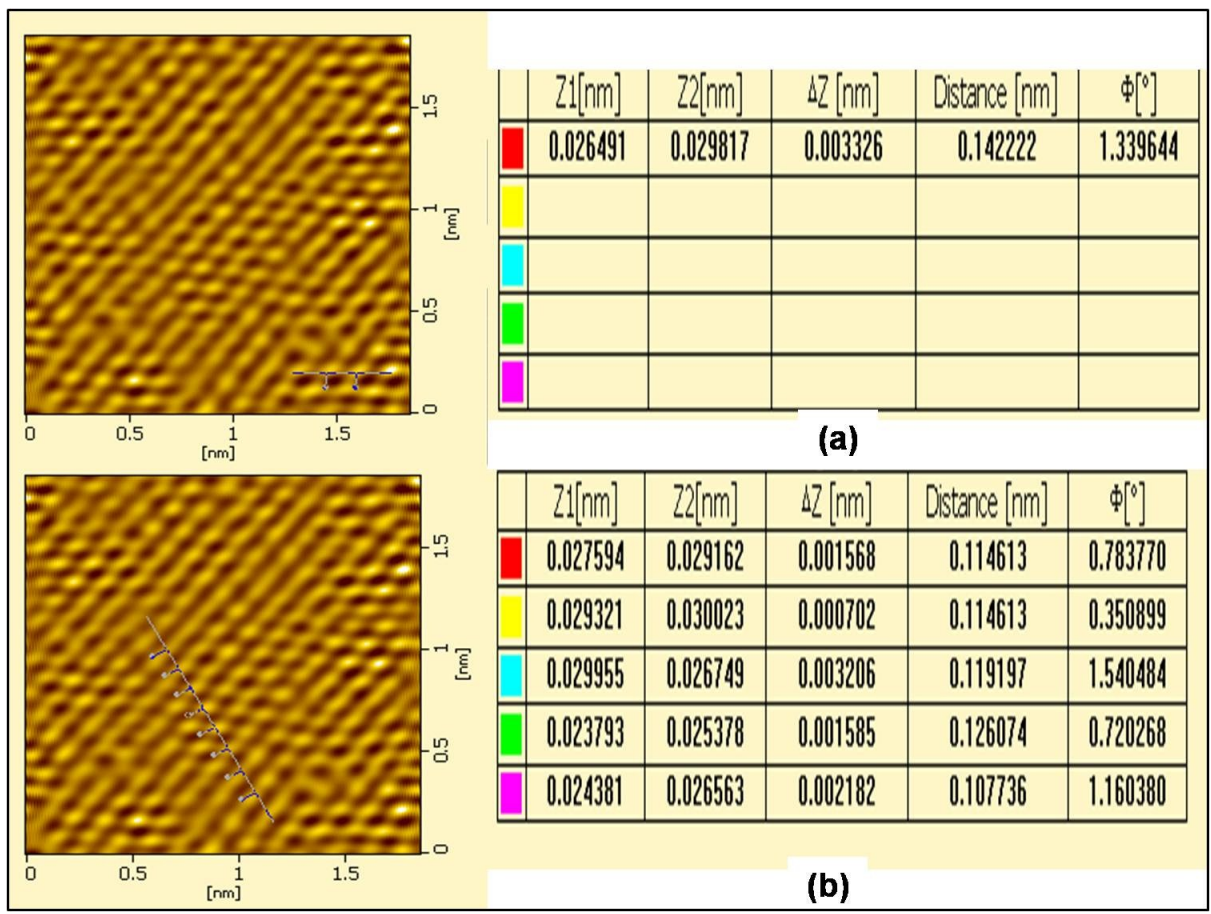

**Figure 7B.** Atomic resolution images showing the presence of (a) graphene and (b) collagen.

*3.7. Confocal Microscopy*

Confocal fluorescence microscopy (CFM) is an established three-dimensional imaging technique in biological and medical sciences. However, its application in colloidal and polymeric systems is just beginning to be explored; basically, it uses laser light to excite fluorophores, either present inherently in the system or added externally and forms an image from the emitted light. Though many biopolymers are often fluorescently tagged and imaged with CFM it has rarely been done with C dispersions [29-31]. Imaged in the liquid state, the auto-fluorescence of triple helix of C in the acetic acid-water system is seen both in green and red (Fig.7A a-b). Whenever, a sparingly soluble biopolymer/polymer is used as a reactant/catalyst in a given reaction, the microenvironment differs from that of solution chemistry, it is more reactive because of the presence of dangling/broken/sacrificial bonds and self assembles to thermodynamically stable clusters. Previous studies involving C have revealed that the triple-helices are arranged in a hexagonal or quasi-hexagonal array in cross-section, in both the gap and overlap regions [32, 33]. It is this structural similarity with GR that leads to favorable interaction and mutual co-existence. GR flakes in water shows more auto-fluorescence as compared to C, and the zig-zag protruding edges are clearly visible (Fig.7A c-d). When GR is incubated with C, (molecular weight of 300kDa, and isoelectric point of 8.26) at pH 3, it is positively charged and is expected to have a hydration layer around it, which would initially attack the zig-zag edges of the GR flakes in water.

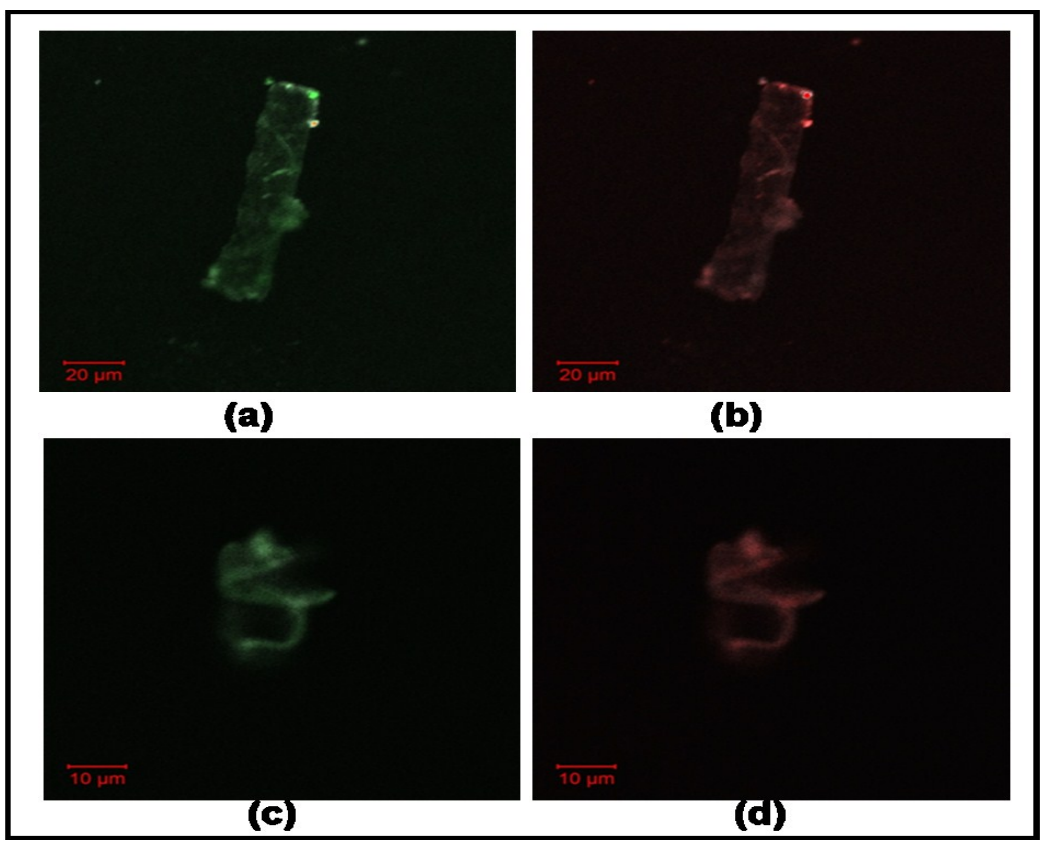

**Figure 8A.** Confocal micrographs of C and GR dispersions showing auto-flourescence. Images (a-b) are C triple helix, (c-d) graphitic stacks in water

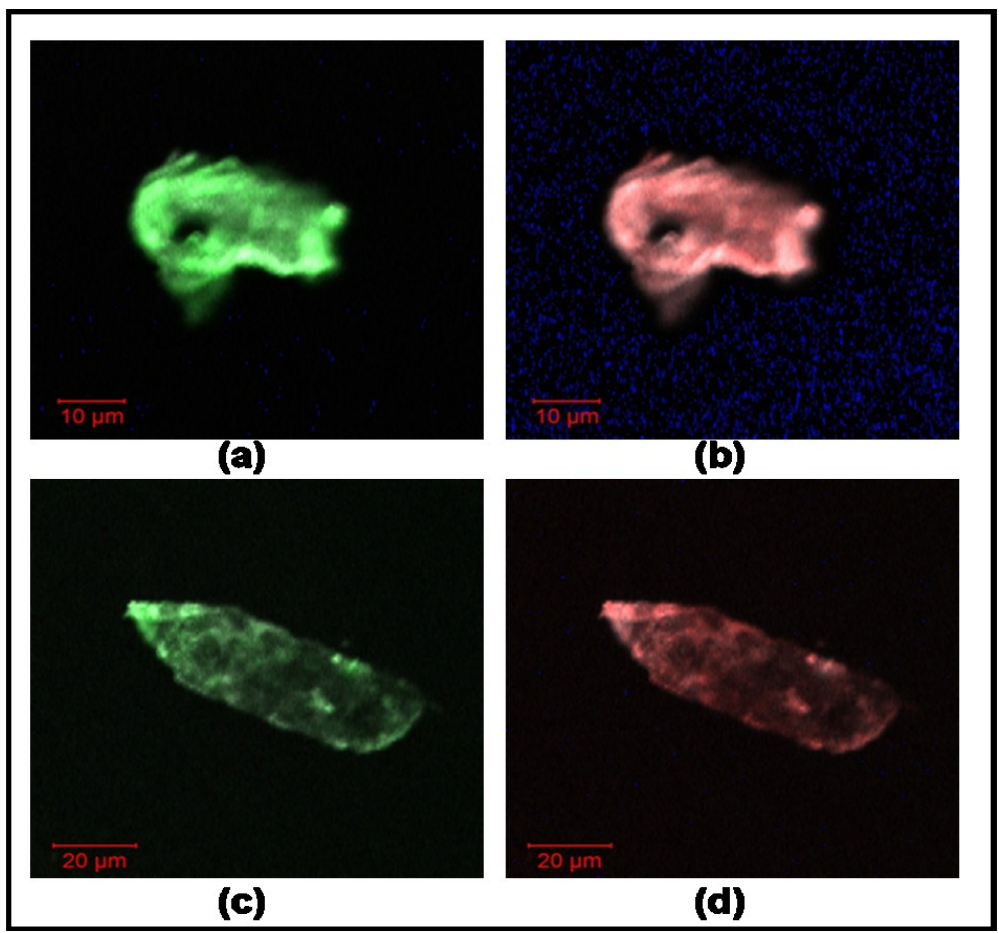

**Figure 8B.** Confocal micrographs of (a-b) GC1 and (c-d) GC7. The structure is more open after 1 day incubation; the graphitic planes exfoliate and reassemble with time

It is noteworthy that the auto-fluorescence of GC1 is more than that of C and GR, because of the additive fluorescence of both. The shape of the visualized clusters is clearly, a combination of C and GW (Fig. 7B a-b). To produce colloidal G, the π-stacking has to be disturbed, previous reports have shown that molecules having surface energy closely matching the surface energy of GR can exfoliate it. It is this energy matching provided by the sacrificial bonds and structural similarity of the two reactants that helps form colloidal G. The kinetics of this process after seven days can be explained from the captured confocal images of GC7 (Fig. 7B c-d). It definitely appears more transparent than GC1, chances are that either the collagen is peeling off graphitic layers with time or that the composite structures are existing in a stable sheet like open structure.

*3.8. Thermal Conductivity and viscosity measurements*

Recent work on nanofluid heat transfer has shown that carbon nanotube (CNT) based nanofluids exhibit excellent thermal properties [34]. With the discovery of graphene, studies have been initiated to measure the heat transport properties of graphene nanofluids [35]. A fluid containing graphene is expected to increase the thermal conductivity tremendously which can revolutionize biomedical therapeutics like controlled drug-delivery assisted thermal ablation of internal carcinoma cells. No work has still been reported where protein assisted directly exfoliated graphene colloids have been tested as potential thermo-fluids for hyperthermia treatments, making our work significant in this regard. The thermal conductivity measurements on colloidal graphene showed interesting results; all comparison were made w.r.t. distilled water as the base medium. Sets of data were recorded by varying the temperature periodically from 283K (10$^o$C) to 313K (40$^o$C) at an interval of 5K after 12, 17 and 22 days respectively. Figures 8 (a & b) show the increase in thermal conductivity of the sonicated sample with rising temperature followed by the percentage enhancement while in figures 8 (c & d) the values for the unsonicated sample has been shown. The % enhancement was calculated from the equation

$$\left(\frac{K_{GC} - K_0}{K_0}\right) \times 100$$

where $K_{GC}$ is the thermal conductivity of colloidal graphene and $K_0$ is the thermal conductivity of distilled water. Considering the fact that both the samples had the same volume fraction of GC composites (~0.00005), it was observed that after sonication the enhancement was higher compared to the unsonicated sample. The highest recorded enhancement after sonication was 16.8 after 22 days followed by 15.9 after 17 days and 15.1 after 12 days; all values recorded at 313K (40°C). The increase in conductivity with temperature followed a linear trend for both the samples; however it was also observed that the increase was only marginal after 12 and 17 days but increased more after 22 days. The values remained constant at this value when the experiments were conducted a 4th and 5th time respectively after a month. The increase and final stabilization of the thermal conductivity of the fluid can be treated as qualitative evidence for the transient exfoliation of GR by C to form G. The fact that in the early experiments the thermal conductivity increments are lower than the later experiments implies that GR (thermal conductivity lower than that of G) is yet to be completely transformed to G. With time, the increase in thermal conductivity rises, implying gradual transformation of the remaining GR to G. At the end of the transformation process, wherein the GR has been completely exfoliated to G, the thermal conductivity value reaches a stand-still and further increment with time is not possible for a particular sample. This final value is the highest thermal conductivity that can be obtained for a fixed sample with a fixed concentration of GR. The fact that with a very low volume fraction of GC composites in dispersion, (G is lesser (not estimated) )we get nearly 17% rise in conductivity, gives ample scope for further studies with the protein chemistry and synthesizing better graphene colloids.

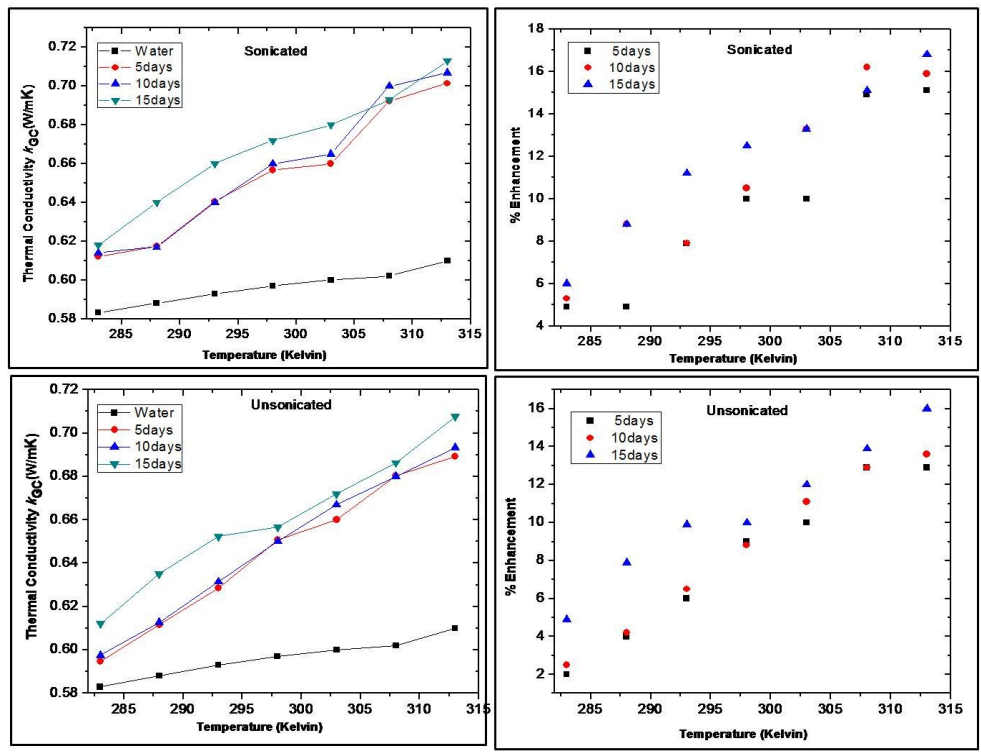

**Figure 9A. (a)** thermal conductivity of colloidal graphene after sonication, (b) corresponding % enhancement, (c) thermal conductivity of colloidal graphene unsonicated and (d) corresponding % enhancement

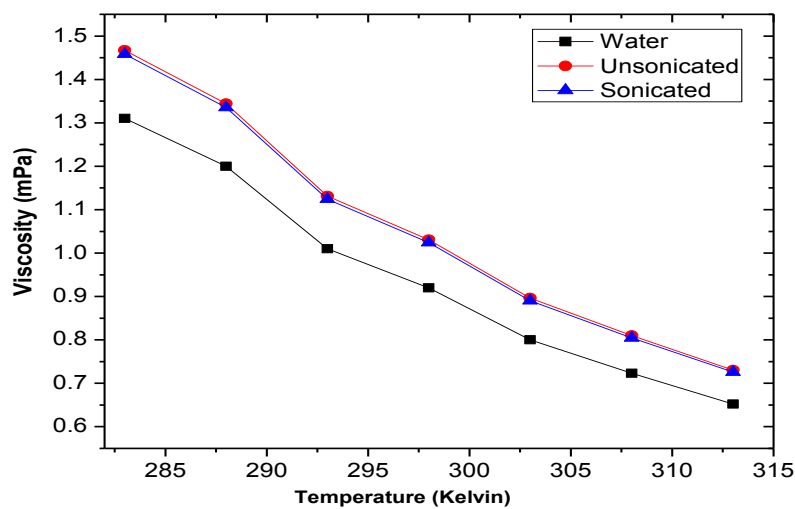

**Figure 9B.** Viscosity of water, unsonicated and sonicated samples

The viscosity data for both the fluids is just 10-15% higher than the base fluid viscosity. Viscosity measurements were performed using an Ostwald's viscometer, with water as the reference. The values of viscosity were cross-checked using an Anton Paar automated micro-viscometer. The values at a given temperature were found to be within 5% of each other. The low viscosity of the fluids is of great utility if they are used as carrier fluids in biomedical applications such as targeted drug delivery and intra-tumor injections for Laser assisted hyperthermia.

## 4. Conclusions

The ability to produce stable suspensions of G in C dispersion, in bulk, is itself an achievement. The main challenge that remains, even today, in G synthesis, is to produce a large enough volume of G safely and in a cost efficient manner. Time kinetics study of C-GR interaction poses a question if it is too early to predict that the C and GR interaction is thermodynamically very favorable since the confocal micrographs reveal extensive interaction with time and the XPS data confirm sequential changes in the chemical environment around the C1s. Nothing else can explain the visual observation of greyish-black coloration of the colloidal dispersion, its stability and zero additional energy input. The dramatic rise in PL intensity and the development of the new peak in fluorescence signify the disturbance of the tertiary structure of C. Hybrid materials with novel and enhanced properties are being extensively studied nowadays but this study is a class apart as the process is also green with zero energy involvement. The AFM study shows the presence of both collagen and graphene, thereby providing evidence that the biological properties of collagen remains unaltered during this process. This provides scope for future applications of C-G as possible stable bio-compatible composites. The high enhancement in thermal conductivity of C-G colloids at low volume fraction of G, with minute increase in the viscosity qualifies the fluid as a novel bio-nanofluid for possible applications in intra-tumor injections for laser assisted hyperthermia, bio-compatible drug carrier for magneto-hyperthermia of deep seated carcinoma tissues and for targeted drug-delivery to internal organs. The high stability of the samples gives ample scope of increasing the graphene content in the colloidal composite by playing with the protein conformation.

**Acknowledgements**

The authors acknowledge the contributions of N R Bandyopadhyay and Subhabrata Chakraborty for TEM characterization, J K Singh and Archana Pandey for Raman spectroscopy, Soumen Mondal for XRD, R K Sahu and Y N Singhbabu for XPS , Sujoy Dey for confocal studies and A.K.Pramanik for AFM. S Bhattacharya also acknowledges the research support of CSIR-Senior Research Fellowship, Govt. of India.

**LEGENDS**

**Figures**

Figure 1. GW, GC1 and GC7 dispersions

Figure 2. SERS spectra of C, GC1 and GC7. The 2D (2727cm$^{-1}$) peak in GC1 vanishes in GC7 signifying interaction. The control spectra are shown in the inset

Figure 3. XRD peaks of C, GC1 and GC7, the controls shown in the inset. Highest intensity peak was recorded at 26.67° corresponding to (002) reflection

Figure 4. TEM micrographs of GC 7 dispersions (a) - (d) showing the seamless co-existence of C and G

Figure 5. (a) FL spectra of C, GC1 and GC7 showing conformational changes in the protein structure (excitation wavelength ~ 280nm, emission range ~ 290-450 nm) (b) PL spectra of the samples proving opening up of graphitic stacks (excitation wavelength ~ 280nm, emission range ~ 290-450 nm). The decrease in FL and increase in PL correspond well with the gradual opening of the system

Figure 6. XPS spectra of (a) GR, (b) C, (c) GC1 and (d) GC7 dispersions suggesting interaction between C and GR

Figure 7A. Non-contact AFM images of GC composites on glass (focus area equal to 4μm x 4μm) in different modes (a) topography (b) frictional force and (c) deflection, (d) a topographical image of the surface showing a bigger area of focus 1000 μm x 1000 μm.

Figure 7B. Atomic resolution images showing the presence of (a) graphene and (b) collagen .

Figure 8A. Confocal micrographs of C and GW dispersions showing auto-flourescence. Images (a-b) are C triple helix, (c-d) graphitic stacks in water

Figure 8B. Confocal micrographs of (a-b) GC1 and (c-d) GC7. The structure is more open after 1 day incubation; the graphitic planes exfoliate and reassemble with time

Figure 9B. Viscosity of water, unsonicated and sonicated samples

Figure 9A. (a) thermal conductivity of colloidal graphene after sonication, (b) corresponding % enhancement, (c) thermal conductivity of colloidal graphene unsonicated and (d) corresponding % enhancement

**Tables**

Table1. Detailed Raman peak positions of the analyzed samples corresponding to D, G and 2D bands along with the intensity ratios ($I_D/I_G$ & $I_{2D}/I_G$) signifying interaction

Table 2. XRD parameters that have undergone change as a result of C interaction

Table 3. Spectral parameters showing B.E., peak width and % concentration of deconvoluted peaks corresponding to the C1s spectra of the samples, signifying different functional species bonded with parental carbon